**Improving Testability and Reuse by Transitioning to Functional Programming**

Morgan C. Benton and Nicole M. Radziwill

*Declarative styles such as functional programming (FP) are rapidly gaining ground on their imperative cousins, including procedural and object-oriented programming. The shift is subtle because it is happening within the context of multiparadigm programming languages such as JavaScript. FP is better suited to modern processes like test-driven development (TDD), and architectures like massively parallel, cloud-based computing. This article describes the technical details that characterize the shift from imperative to FP and implications for software quality management, particularly reuse and testability.*

## INTRODUCTION

Managing the development and maintenance of software is notoriously difficult and constantly evolving (Highsmith 2013). Software process improvement (SPI) approaches, such as the Capability Maturity Model® Integration (CMMI) (CMMI Product Team 2010), have helped software development teams focus on continuous improvement, allowing them to evaluate and incorporate quickly-evolving strategies and standards, such as agile development processes like Scrum (Łukasiewicz and Miler 2012).

For software in particular, it is important to build quality in at the design stage rather than relying on inspection. Quality cost models applied to software development (for example, Gack 2011, 8) reveal that "judiciously chosen increases in prevention and appraisal will reduce rework and lead to an overall increase in the portion of total effort that is value added... [meaning] the application of best practices at the right time in the development life cycle." Choosing the programming paradigm that best suits the intended deployment environment can optimize value-added activities and result in fewer lines of code and reduced coupling to other components (Singh and Saha 2007 **AUTHOR: REF. LIST SAYS 2009. WHICH IS CORRECT?).** It can also influence the ability to implement reusable components and optimize future potential reusability (Ranjan and Tripathi 2009).

This article identifies and describes a potentially disruptive shift in the way software is being designed and built, and suggests strategies that will allow the software development managers and teams who choose it make a smoother transition to this new model.

**From Imperative to Declarative**

> "There are these two young fish swimming along and they happen to meet an older fish swimming the other way, who nods at them and says 'Morning, boys. How's the water?' And the two young fish swim on for a bit, and then eventually one of them looks over at the other and goes, 'What the hell is water?'"
>
> *--David Foster Wallace, 2005*

There are many different paradigms for creating software (Van Roy 2009). The essential feature of a paradigm is that it is *fundamental* to the way programmers think about the design and implementation of a program -- so fundamental, in fact, that many or most coders will not even be aware that they follow a paradigm. Like air to humans, or water to fish, the paradigm is ubiquitous, invisible, and taken for granted, yet crucial to survival. A coder's paradigm influences nearly every action and decision that is made in the course of development, from choice of a development environment, to architecture, to how one conceptualizes the processes that software automates, and certainly how one communicates about it. It is difficult to overstate how important it is that managers and developers understand not only that their efforts are encapsulated in a paradigmatic worldview, but also that other world views exist, and, furthermore, that they are capable of moving between and even mixing them.

People identify passionately, even dogmatically, with their world views, and changes can be unsettling, uncomfortable, and even distressing. In the scientific world, paradigm shifts happen when growing evidence supports a model of the world that contradicts, supersedes, or otherwise conflicts with the current dominant model. Examples include the shift from a geocentric to heliocentric model of the solar system, or the shift from Newtonian to relativistic physics (Kuhn 2012). While it is tempting to get bogged down in fun, esoteric discussions like, "Does the shift from imperative to FP rise to the level of paradigm shift?" or "Are programming paradigms really 'paradigms' in the true Kuhnian sense of the word?" the point made here is that *the current shift in programming practice has the potential to be disruptive*, and regardless of its definition or magnitude, it is something software quality professionals should pay close attention to and prepare for.

**PROGRAMMING PARADIGMS**

Programming paradigms can be put into a simplified hierarchy (see Figure 1). This is a very simplified representation of the world of programming paradigms. For a much more thorough, nuanced introduction to the subject, the authors recommend Highsmith (2013), who describes roughly 30 distinct paradigms and provides concrete criteria for how his taxonomy was created. What is significant is that since its inception in the 1940s, imperative programming has dominated software development, but now that focus is shifting to the declarative branch of the tree. This is a tipping point in the maturation of programming practice.

None of these paradigms are new. For example, functional programming (FP), which is the focus of most of this article, has its roots in the lambda calculus developed in the 1930s by Alonzo Church, a teacher and collaborator of Alan Turing -- which is to say, FP predates the first computer. Imperative programming, on the other hand, is tightly coupled conceptually with most computer hardware, which is designed to follow a set of instructions. What is new and noteworthy is a shift from imperative programming to FP as the dominant paradigm.

**A Shift from "How" to "What"**

Put simply, the current shift represents a change in focus from the question of "How do I make the computer do what I want it to do?" to "What is it that I want the computer to do?" This may be easiest to illustrate with example code. This example uses the FizzBuzz Problem (Ghory 2007), which is:

> *Write a program that prints the numbers from 1 to 100. But for multiples of three print "Fizz" instead of the number and for the multiples of five print "Buzz." For numbers which are multiples of both three and five print "FizzBuzz."*

JavaScript is used because it supports both imperative and declarative syntax, and also because it is one of the most widely used programming languages in the world today. Please note that the functional FizzBuzz example is not pure FP, as it does not isolate all side effects. It is essentially declarative, however, and sufficient to illustrate some reasons FP is gaining popularity.

These blocks of code produce identical output but approach the solution from very different viewpoints. The imperative code describes *how* to produce the output very carefully in a step-by-step fashion. It tells the computer sequentially to:

1. Start with the number 1.
2. If it is divisible by 3 and 5, output "FizzBuzz" and then move on to the next number.
3. If it is divisible by 3, output "Fizz" and then move on to the next number.
4. If it is divisible by 5, output "Buzz" and then move on to the next number.
5. Otherwise, output just the number.
6. If the next number is less than 101, repeat steps 2-5, otherwise finish.

These instructions are an indivisible unit. There is no way to write a test that can single out one particular instruction without testing them all. In a more "real world" scenario, if there is a problem somewhere in these instructions, it may be very difficult to figure out where the bug is.

The FP example first creates four functions that describe what functionality is desired. The first three (fizz, buzz, and fibu) take a single parameter and convert it into a specified string if it meets certain criteria, and otherwise pass along the original value. The next function, log, simply writes its input to the console. It's not until the fifth function, fizzbuzz, that really interesting things start to happen. The fizzbuzz function is a composition of all the earlier functions. To translate what it does into English, it says, "For each element of an array, pipe it through these four functions from right to left." Next, it says, provide an array filled with the numbers from 1 (inclusive) to 101 (exclusive). Finally, this array of numbers is piped through the fizzbuzz function.

Although it may not be readily apparent from such a small example, the code on the right has some properties that make it a better solution for solving modern software problems, namely:

1. **Testability**: Almost every line of the program describes a testable function.
2. **Reusability**: Assuming each function encapsulates some useful bit of functionality, each could easily be reused to compose other solutions.
3. **Functional independence**: The internal implementation of each function is unimportant; by using _.map() one is not bound to a sequential, for-loop-style iteration over the collection of numbers; that processing may be parallelized or run in whichever way is most efficient.

The authors would like to emphasize the first key difference: if there was a bug in this code, it could easily be traced to the individual function responsible. This article argues that these differences are highly significant and drive a paradigm shift in programming that will bring great benefits to teams that exploit them sooner rather than later. The next section of this article will examine each of these differences in more detail, but first the question of "Why now?" is addressed.

**Why FP? Why Now?**

As mentioned earlier, the concepts underlying FP are older than computers. Lisp, the first FP language, has been around since the 1950s. Why is it that these concepts are only now becoming mainstream? The answer is that the computing ecosystem has not been able to support widespread use of FP until very recently. Three elements contribute to this new FP-friendly ecosystem.

The first element of the new ecosystem is the ability to make use of encapsulated best practices. Earlier it was stated that one of the strengths of FP is that it allows one to focus on the *what* instead of the *how*. It is crucial to understand that the ability to focus on the *what* question presupposes that the *how* question has already been answered. Until recently, however, the coding world was still very much preoccupied with figuring out how to make computers do things. The _.map() function used previously takes its argument and applies it to every element in an array. The coder using _.map() may not know or care what the underlying implementation of that function is, but what makes using it possible is that someone has already solved that problem. Indeed, the entire FP example relies on an external library (Ramda) that encapsulates previously solved problems and frees the developer to focus only on the problem at hand. As a community, developers have begun to take the "Don't Repeat Yourself" (DRY) principle seriously and leverage their collective experience in support of one another.

The second element of the ecosystem driving adoption of FP is the culture of sharing fostered by the open source software (OSS) movement. Not just individual developers, but also companies of all sizes have embraced the idea that everyone is better off if they share their code. The degree to which innovation is catalyzed by having a rich array of high-quality software tools at one's fingertips is hard to calculate. Whether it is Google's Angular framework, Facebook's React, or any of the thousands of high-quality packages available at NodeJS, Ramda, or jQuery, these pieces of code are made available because one recognizes that many others will take these building blocks and make wonderful things that will enrich people's lives more than any money they could have made by trying to sell the code.

The third element of the ecosystem catalyzing FP is a collective awareness of the cost of bugs and the necessity of earning trust in one's coding ability by writing tests. The OSS movement could not thrive without thousands of developers all over the world working independently to contribute pieces of the puzzle. Those developers would not be able to trust each other unless they had a mechanism for demonstrating that their code was not going to hurt people. This mechanism is testing. As testing has increasingly become an integral part of even the smallest projects, developers are increasingly finding ways to make every line of code that they write testable. Since testability is one of the core properties of FP, it makes sense that the community would naturally gravitate toward a paradigm that supports how it operates.

In summary, the reason FP is taking off now is that programmers, as one global community of practice, are growing up. They have suffered through many decades of buggy, hard-to-maintain code, and they have finally figured out how to build an infrastructure that will make following best practices the default course of action. They are finally reaching a stage when they can truly spend the majority of their time thinking about *what* they want to build, not *how* they are going to build it.

**A CLOSER LOOK AT FP**

Let's take a closer look at the differences pointed out previously between the imperative coding and FP implementations of FizzBuzz. Each of these differences has profound implications for software quality that together make a strong argument for why quality-minded managers should begin to think about developing and incorporating more capacity for FP in their software development teams, if they haven't done so already.

**Testability**

If a team is not already engaged in test-driven development (TDD), it's likely they soon will be. In a 2010 study of mainstream adoption of agile methods, 84 percent of the teams surveyed indicated that they were already using TDD or planned to do so (West et al. 2010). Code written under the FP paradigm is testable by design, and as such is a driver for greater FP adoption moving forward. It should also be noted that this is likely to be a self-reinforcing loop; greater adoption of TDD pushes teams toward paradigms like FP that yield more testable code, and writing more testable code in turn facilitates teams moving toward TDD.

Examining the imperative version of the FizzBuzz code in Figure 2, there appears to be no way to test that code other than as one monolithic entity. The functional version, however, easily lends itself to testing, as Figure 3 illustrates.

While this kind of testing is certainly overkill for this little FizzBuzz program, it demonstrates how easily a TDD approach can be adopted when working with FP implementations. Even better, the use of Hindley-Milner type signatures (Milner 1978) will even suggest what tests to write, or how to write the type signature based on the tests.

**Reusability**

One of the chief goals of object-oriented programming (OOP) has been to foster clean design of code that would allow that code to be reused in future projects. The popularity of discussion on this topic on StackExchange indicates that OOP's success in this area is somewhat in question. However, FP fosters true abstraction of functionality through practice at writing Hindley-Milner type signatures. If written correctly, these type signatures allow developers to search for functionally equivalent code through search engines like Hoogle. While Hoogle is specifically for the Haskell programming language, it seems plausible that a language-specific version of any function that could be described with a Hindley-Milner type signature could likely be imported directly into a code editor.

While a full discussion and tutorial on Hindley-Milner type signatures is beyond the scope of this article, a brief description seems in order. The type signatures are typically entered into the code as a comment directly above the function being described. They serve as another means of documenting the code, and can be quite expressive once the syntax is learned. For example:

```
// capitalize :: String -> String
var capitalize = s => toUpperCase(head(s)) + toLowerCase(tail(s));
```

Here there is a function called "capitalize" that takes a string as input and returns a string. One could find all related functions by searching for those that take a string as input and return a string as output. Here's one more example:

```
// strLength :: String -> Number
var strLength = s => s.length;
```

While this function seems a little pointless given that it reproduces functionality built into the programming language, one can at least see that there is a function that takes a string as input and returns a number as output.

In the FizzBuzz example, there are several functions like this one:

```
// fizz :: String -> String
var fizz = x => (0 === x%3) ? 'Fizz' : x;
```

While it may be difficult to imagine another scenario in which one might want to reuse this particular bit of functionality, the broader point is that there is a simple, easily intelligible component that transforms string input according to a specific set of rules. If one wanted to develop a new game called FizzBuzzBlam, it would be trivial to incorporate the fizz function into this new context.

At a larger level of abstraction, FP holds promise for fostering better large-grained software reuse. Witman and Ryan (2010, 42, 147) describe fine- to medium-grained reuse as "ranging from objects, subroutines, and components through software product lines," whereas large-grained reuse would entail entire systems of software products and/or hardware infrastructure. "Good system architecture" was credited as one of the key factors supporting large-grained reuse. As demonstrated previously, FP is driven by flexible, testable architecture, and conscientious use of Hindley-Milner type signatures may result in search engines that could quickly analyze the suitability of a system of software packages for reuse in other contexts.

**Functionally Independent**

The third key difference between these imperative and functional FizzBuzz examples has to do with the functional independence of the lines of code that were written. This loose coupling, while not unique to FP, is a top priority and is one of the things taken for granted in the FP paradigm. One thing to notice about these two implementations is that the imperative FizzBuzz essentially cannot be broken down into any simpler parts -- it is a monolithic block of code. In contrast, the declarative FizzBuzz is built incrementally from very small, interchangeable parts. This has two important ramifications.

First, this impacts how labor is divided among members of a programming team. Whereas the imperative FizzBuzz would almost certainly have to be built by a single programmer, the different pieces of the FP FizzBuzz could be assigned to different members of the team. More developers working on smaller, simpler pieces means a faster, more coherent process. If using TDD, the pieces that need to be built would become apparent when writing the tests.

Second, functional independence means that the internal implementation of the pieces becomes mostly irrelevant from the perspective of the composed program. A function designed to process an array could do so sequentially or in parallel, could send the array off to a Web service for processing, or anything else it wanted as long as the output meets functional requirements. Different implementations might be used in different contexts. This is very important when the diversity of modern computing environments is taken into account. This issue is revisited in the next section.

**DISCUSSION**

In this section, four issues are discussed: suggestions of contexts that are amenable to FP, the question of FP vs. FP languages, whether the shift toward FP is intentional, and how software quality managers should prepare for this change.

**Contexts Amenable to FP**

FP languages naturally align with modern deployment environments. Although alluded to throughout this article, the specific ways that FP is a good fit for various modern programming contexts have not been explicitly addressed. This section rectifies that. Each scenario discusses why FP may be a better fit than the current imperative approach for several contexts.

**Internet of Things**

Internet of things (IoT) refers to the coming era in which "visions [of] 50 billion Internet-connected devices become a reality" (Baresi et al. 2015, 6). This is the world in which everything from a coffee maker, to a pair of shoes, to a child's backpack is connected to the Internet. These things communicate with, or can be controlled by, applications on one's phone, or through new interfaces that haven't been developed. In their introduction to *IEEE Internet Computing's* special issue on IoT, Baresi, Mottola, and Dustdar (2015) describe the challenge of knitting together a multidisciplinary patchwork of code at the networking, middleware, and application layers.

In this kind of environment, it would be a blessing for a developer to be able to focus on *what* and not *how* to build these devices and applications at every layer, a capability that is enabled by FP. Given the very limited storage space in these embedded systems, FP's immutable types would seem to provide maximum functionality with a minimum of the memory overhead needed to store information about the state of systems.

**Big Data Analysis and Cloud Computing**

Frequently, "big data" sets (of 1 TB or larger) reside on multiple servers that may be spread throughout the cloud. Analysis involves millions of mathematical calculations, many of which are most efficiently processed in parallel. Mathematical reasoning to assure the statistical soundness of the results is of prime importance. Data scientists tend not to be very interested in how this analysis happens at the algorithmic or procedural level, and prefer to focus on the big picture, that is, the questions that can be answered with these large data sets. At the very least, a functional interface to the data (for example, using the R or Python programming languages) would be of interest to researchers and analysts in this field.

**Agile Software Development**

The FizzBuzz example shows FP's capacity for breaking programs up into very small, testable chunks. This is an ideal fit for the rapid iterations and parallel work of a modern agile team. High testability means less frustration when it comes time to integrate the work of multiple developers together in the repository. It also means that when it comes time for profiling and refactoring, it will be much easier to pinpoint the functions that are causing bottlenecks and need to be reimplemented.

**Functional Programming vs. Functional Programming Languages**

One issue that may be confusing is that FP is typically associated with certain programming languages that were designed primarily for FP, such as Haskell, Scheme, Standard ML, or Scala. In fact, one can write functional-style code with any language that supports the following:

- First-class and higher-order functions
- Recursion
- Anonymous functions and classes

Consequently, JavaScript, PHP, C#, and Python can all be used to write functional code. As an example, Figure 4 demonstrates an FP version of the FizzBuzz problem written in C#. Many recent, popular application development frameworks such as AngularJS and Laravel have adopted a functional style. An interesting question is whether this represents an intentional shift in style, or if coding practices have naturally evolved in this direction.

**Is the Shift to FP Intentional?**

The FP paradigm seems to address many of the issues of modern software development. Indeed, some of the updates to modern languages have been to add features to support FP. For example, one of the additions to PHP7, which was officially released in December 2015, was support for anonymous classes (*The PHP Manual* 2015). Similarly, ECMAScript 6 (that is, JavaScript) added a more expressive closure syntax, which was used in the aforementioned functional FizzBuzz example to make the code more terse and easier to read (Engelschall 2015). Frameworks such as D3, Angular, and Laravel seem to be guiding developers in the direction of writing more functional code. Primarily functional languages such as R are growing in popularity.

That being said, the words "functional programming" do not appear in the official release notes, nor often in the documentation for these frameworks. There are plenty of websites that provide tutorials on how to write functional code in any of these languages, and some developers are explicit about their preference for FP, but by and large, the shift to a more functional style in various places seems driven purely by the challenges and pragmatics of modern software development. It is almost as if the programming community is slowly learning via trial and error to recreate the lambda calculus that Alonzo Church developed more than 75 years ago.

Regardless of whether the shift is being made subconsciously or intentionally, it is important to recognize that the shift *is* happening. As such, developers and managers would do well to take an explicit, structured approach to learning this new way of thinking. Such training would be preferable to the current mode of shifting, which is slow, subconscious, and unstructured.

**How Should Development Managers Prepare for FP?**

Even though testing can cost from 30 to 50 percent of the total development budget, the cost of internal and external failures is even greater (Cohen 2004). Though development managers may recognize the benefits of test-driven development and the enhanced potential for code reuse, deciding to recommend a new approach like FP can be risky. They may wish to gather tangible evidence that an approach will yield benefits in a particular problem domain before supporting a change. Fortunately, there are models for conducting controlled experiments to establish whether a new programming or testing approach affects software quality (Juristo and Moreno 2013). A Latin Square design, for example, can be used to evaluate a small-scale programming task to determine whether a choice of paradigm (FP, procedural, or object-oriented) influences response variables such as time to code completion, total development time, defect density, or code coverage, while controlling for the variance introduced by different developers or teams working on different days (Lauterbach and Randall 1989). **Hilton (2009) AUTHOR: THIS IS NOT ON REF. LIST. PLEASE ADD** recommends that the response variables should be design-based (for example, cohesion, coupling, or complexity measures). Like with any empirical study, there will be threats to validity that development managers and architects should take into consideration: developers may be more motivated to deliver higher-quality software for the programming approach they are most interested in using, experimental tasks may be easier than production tasks, and there may be learning curves associated with new programming approaches for some developers (Bhat and Nagappan 2006).

**CONCLUSION**

Whether it is because of natural evolution or intentional steering in that direction, the state of the art for the software development community is shifting toward increased use of the FP paradigm. While it will take time for individual developers to fully comprehend and internalize this new style of coding, the benefits to be gained are substantial, particularly for supporting reuse and enhanced testability for IoT and other cloud-based applications. Given its mathematical properties, and a philosophy that goes hand-in-hand with test-driven development, FP will be a new friend to quality-minded managers and coders alike. It will likely open up possibilities for new, more informative metrics on team productivity and code quality, and yield insights that will drive software process innovation efforts. Managers and coders involved in software development are strongly encouraged to begin developing their individual and team capacities in this area.

**REFERENCES**

Baresi, L., L. Mottola, and S. Dustdar. 2015. Building software for the Internet of things. *IEEE Internet Computing* 2:6-8.

Bhat, T., and N. Nagappan. 2006. Evaluating the efficacy of test-driven development: Industrial case studies. In *Proceedings of the 2006 ACM/IEEE International Symposium on Empirical Software Engineering*, 356-363. New York: ACM.

CMMI Product Team. 2010. CMMI® for Development, version 1.3: Improving processes for developing better products and services (CMU/SEI-2010-TR-033). Pittsburgh: Software Engineering Institute.

Cohen, M. B. 2004. Designing test suites for software interaction testing. PhD diss., University of Auckland.

Engelschall, R. 2015. Arrow functions: Expression bodies. ECMAScript 6: New Features: Overview and Comparison. Available at: http://es6-features.org/#ExpressionBodies.

Gack, G. 2011. Powerful metrics: Software cost of quality+ defect containment. *Software Quality Professional* 13, no. 2.

Ghory, I. 2007. Using FizzBuzz to find developers who Grok coding. *Imran on Tech*. Available at: http://bit.ly/1e7Mz4q

Highsmith, J. 2013. *Adaptive software development: A collaborative approach to managing complex systems*. Boston: Addison-Wesley.

Juristo, N., and A. M. Moreno. 2013. *Basics of software engineering experimentation*. New York: Springer Science & Business Media.

Kuhn, T. S. 2012. *The structure of scientific revolutions: 50th anniversary edition*. Chicago: University of Chicago Press.

Lauterbach, L., and W. Randall. 1989. Experimental evaluation of six test techniques. In *Proceedings of the Fourth Annual Conference on Computer Assurance. COMPASS'89, Systems Integrity, Software Safety and Process Security,* 36-41. New York: IEEE.

Łukasiewicz, K., and J. Miler. 2012. Improving agility and discipline of software development with the Scrum and CMMI. *IET Software* 6, no. 5:416-422.

Milner, R. 1978. A theory of type polymorphism in programming. *Journal of Computer and System Sciences* 17, no. 3:348-375.

*The PHP Manual*. 2015. New features in PHP 7. Available at: http://php.net/manual/en/migration70.new-features.php

Radziwill, N. M. 2007. Quality management in astronomical software and data systems. In *Astronomical Data Analysis Software and Systems* XVI, vol. 376:363-372.

Ranjan, D., and A. Kumar Tripathi. 2009. ROSTProM: A reuse-oriented software testing process model. *Software Quality Professional* 11, no. 4.

Singh, Y., and A. Saha. 2009. Assessment of class testability: An empirical study. *Software Quality Professional* 11, no. 3.

Van Roy, P. 2009. Programming paradigms for dummies: What every programmer should know. *New computational paradigms for computer music* 104.

Wallace, D. F. 2005. 2005 Commencement address. Kenyon College, 21 May, Gambier, Ohio.

West, D., T. Grant, M. Gerush, and D. D'silva. 2010. Agile development: Mainstream adoption has changed agility. *Forrester Research* 2:41.

Witman, P. D., and T. Ryan. 2010. Think big for reuse. *Communications of the ACM* 53, no. 1: 142-147.

**BIOGRAPHIES**

**AUTHOR: PLEASE PROVIDE ONE OR TWO PARAGRAPHS FOR EACH AUTHOR. THANKS!**

**Figure 1 A simplified hierarchical view of programming paradigms**

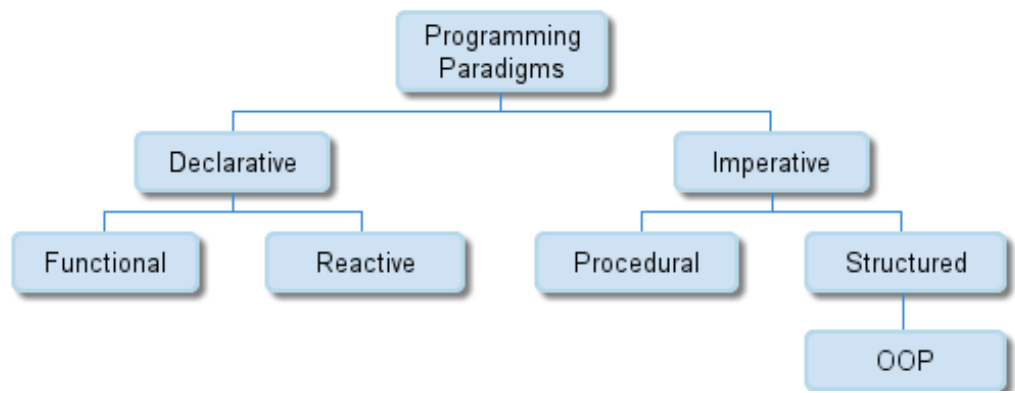

```
// Imperative FizzBuzz                  // Functional FizzBuzz
for (i = 1; i < 101; i += 1) {          require(['ramda'], function(_) {
  if (0 === i%3 + i%5) {                  var fizz = x => 0 === x%3 ? 'Fizz' : x,
    console.log('FizzBuzz');                  buzz = x => 0 === x%5 ? 'Buzz' : x,
    continue;                                 fibu = x => 0 === x%3 + x%5 ? 'FizzBuzz' : x,
  }                                           log  = x => console.log(x),
  if (0 === i%3) {                            fizzbuzz = _.map(_.compose(log, buzz, fizz, fibu)),
    console.log('Fizz');                      numbers  = _.range(1,101);
    continue;                             fizzbuzz(numbers);
  }                                     };
  if (0 === i%5) {
    console.log('Buzz');                // Note: the above code makes use of the RamdaJS library
    continue;                           //       (http://ramdajs.com/), which provides common
  }                                     //       utility functions for writing functional
  console.log(i);                       //       JavaScript, such as _.map() and _.compose()
}
```

**Figure 2 Imperative and declarative solutions to the FizzBuzz problem**

```javascript
// FizzBuzz test using Jasmine (http://jasmine.github.io/)
describe('A FizzBuzz Implementation', function(){
  it('has a fizz function', function(){
    expect(typeof fizz).toBe('function');
  });
  it('that returns "fizz" on multiples of 3', function(){
    var a = fizz(3),
        b = fizz(6),
        c = fizz(123);
    expect(a).toBe('fizz');
    expect(b).toBe('fizz');
    expect(c).toBe('fizz');
  });
  it('and the submitted value otherwise', function(){
    var a = fizz(1),
        b = fizz(7),
        c = fizz('pork chop');
    expect(a).toBe(1);
    expect(b).toBe(7);
    expect(c).toBe('pork chop');
  });
});
```

**Figure 3 Example test for FP FizzBuzz**

```csharp
using System;
using System.Collections.Generic;
using System.Linq;

namespace FPFizzBuzz {
  public class FizzBuzz {
    public static Func<string, bool> isInt = x => x.All(Char.IsDigit);
    public static Func<string, int>  toInt = x => Int16.Parse(x);
    public static Func<string, int, bool> isMult = (x, y) => isInt(x) && 0 == toInt(x) % y;
    public static Func<string, string> fizz = x => isMult(x,  3) ? "Fizz"     : x;
    public static Func<string, string> buzz = x => isMult(x,  5) ? "Buzz"     : x;
    public static Func<string, string> fibu = x => isMult(x, 15) ? "FizzBuzz" : x;
    public static Action<string> log = Console.Write;
    public static Action<int> fizzbuzz = x => log(fizz(buzz(fibu(x.ToString())))  + " ");

    static void Main(string[] args) {
      Enumerable.Range(1, 100).AsParallel().ForAll(x => fizzbuzz(x));
    }
  }
}
```

**Figure 4 Functional implementation of FizzBuzz in C#**